\documentclass[12pt]{article}
\usepackage{a4wide}
\usepackage{graphicx}
\usepackage{xspace}
\newcommand{\eeee}{\ensuremath{e^+e^-\to e^+e^-}\xspace}
\newcommand{\eemm}{\ensuremath{e^+e^-\to\mu^+\mu^-}\xspace}
\newcommand{\eepp}{\ensuremath{e^+e^-\to\pi^+\pi^-}\xspace}

\newcommand{\dd}{\mathrm{d}}
\newcommand{\MeVc}{MeV/$c$\xspace}

\title{Measurement of the $e^+e^-\to\pi^+\pi^-$ cross section 
with the CMD-2 detector in the 370-520 MeV c.m. energy range}

\author{
R.\,R.\,Akhmetshin$^{a}$,
V.\,M.\,Aulchenko$^{a,b}$,
V.\,Sh.\,Banzarov$^{a}$,
L.\,M.\,Barkov$^{a,b}$,  \and
N.\,S.\,Bashtovoy$^{a}$,
A.\,E.\,Bondar$^{a,b}$,
D.\,V.\,Bondarev$^{a,b}$,
A.\,V.\,Bragin$^{a}$,  \and
S.\,K.\,Dhawan$^{d}$,
S.\,I.\,Eidelman$^{a,b}$,
D.\,A.\,Epifanov$^{a}$,
G.\,V.\,Fedotovich$^{a,b}$, \and
N.\,I.\,Gabyshev$^{a}$,
D.\,A.\,Gorbachev$^{a}$,
A.\,A.\,Grebenuk$^{a}$,
D.\,N.\,Grigoriev$^{a,b}$,  \and
\fbox{V.\,W.\,Hughes$^{d}$},
F.\,V.\,Ignatov$^{a}$,
S.\,V.\,Karpov$^{a}$,
V.\,F.\,Kazanin$^{a,b}$, \and
B.\,I.\,Khazin$^{a,b}$,
I.\,A.\,Koop$^{a,b}$,
P.\,P.\,Krokovny$^{a,b}$,
A.\,S.\,Kuzmin$^{a,b}$, \and
I.\,B.\,Logashenko$^{a,c}$,
P.\,A.\,Lukin$^{a,b}$,
A.\,P.\,Lysenko$^{a}$,
K.\,Yu.\,Mikhailov$^{a}$, \and
A.\,I.\,Milshtein$^{a,b}$,
I.\,N.\,Nesterenko$^{a,b}$,
M.\,A.\,Nikulin$^{a}$,
V.\,S.\,Okhapkin$^{a}$, \and
A.\,V.\,Otboev$^{a}$,
E.\,A.\,Perevedentsev$^{a,b}$,
A.\,S.\,Popov$^{a}$,
S.\,I.\,Redin$^{a}$, \and
B.\,L.\,Roberts$^{c}$,
N.\,I.\,Root$^{a}$,
A.\,A.\,Ruban$^{a}$,
N.\,M.\,Ryskulov$^{a}$,
A.\,G.\,Shamov$^{a}$,\and
Yu.\,M.\,Shatunov$^{a}$,
B.\,A.\,Shwartz$^{a,b}$,
A.\,L.\,Sibidanov$^{a}$\/\thanks{e-mail: A.L.Sibidanov@inp.nsk.su},
V.\,A.\,Sidorov$^{a}$,\and
A.\,N.\,Skrinsky$^{a}$,
V.\,P.\,Smakhtin$^{f}$,
I.\,G.\,Snopkov$^{a}$,
E.\,P.\,Solodov$^{a,b}$,\and
\fbox{J.\,A.\,Thompson$^{e}$},
Yu.\,V.\,Yudin$^{a}$,
A.\,S.\,Zaitsev$^{a,b}$,
S.\,G.\,Zverev$^{a}$
}

\date{}
\begin{document}
\maketitle
\begin{center}
  {\it $^a$Budker Institute of Nuclear Physics, 630090,
    Novosibirsk, Russia} \\
  {\it $^b$Novosibirsk State University, 630090,
    Novosibirsk, Russia} \\
  {\it $^c$Boston University, Boston, MA 02215, USA}\\
  {\it $^d$Yale University, New Haven, CT 06511, USA}\\
  {\it $^e$University of Pittsburgh, Pittsburgh, PA 15260, USA}\\
  {\it $^f$Weizmann Institute of Science,  76100, Rehovot, Israel}\\
\end{center}

\begin{abstract}
The cross section of the process $e^+e^-\to\pi^+\pi^-$ has been
measured at the CMD-2 detector in the 370-520 MeV center-of-mass
(c.m.)  energy range. A systematic uncertainty of the measurement is
0.7\%.  Using all CMD-2 data on the pion form factor, the pion
electromagnetic radius was calculated. The cross section of muon pair
production was also determined.
\end{abstract}

\section{Introduction}  
Study of the process $e^+e^- \to \pi^+\pi^-$ provides important
information about the pion electromagnetic form factor, which
describes its internal structure and is predicted by various
theoretical models. Accurate cross section measurement is necessary
for a more precise determination of the hadronic contribution to the
muon anomalous magnetic moment, which is given in perturbation theory
by the dispersion integral~\cite{kinoshita}:\\
\begin{equation}
{a}^{\rm had,LO}_{\mu}=
 \frac{m^2_\mu}{12\pi^3}
\int\limits_{4m^2_\pi}^{\infty} \dd s\: 
\frac{\sigma^{(0)}(e^+e^- \to {\rm hadrons})\:\hat{K}(s)}{s},
\label{eq:dispint}
\end{equation}
where $\hat{K}(s)$ is a function monotonously rising from 0.63 at the
threshold of pion pair production $s=4m^2_{\pi}$ up to 1 at $s \to
\infty$, where $s$ is the total c.m. energy squared. Since the kernel
$\hat{K}(s)/s$ in the integral enhances a contribution from the low
energy region, particularly important is precise knowledge of the
cross section of the process $e^+e^- \to \pi^+\pi^-$, which is
dominating below 1~GeV.  Furthermore, this cross section is needed to
test the relation between the two-pion spectral function in $e^+e^-$
annhilation and $\tau$ decays, based on conserved vector current and
SU(2) symmetry~\cite{davier}.

For analysis we use a data sample of 56~nb$^{-1}$ collected with the
CMD-2 detector at the electron-positron collider VEPP-2M~\cite{VEPP2M}
at 10 c.m. energy points from 370 MeV to 520 MeV.  During the
experiment about one million events were recorded.

A general-purpose cryogenic magnetic detector (CMD-2)~\cite{cmd2}
consists of a tracking system, barrel and endcap electromagnetic
calorimeters based on CsI and BGO crystals, respectively, and a
muon-range system. The tracking system consists of a drift chamber
with jet-like cells and two layers of a proportional Z-chamber inside
a thin superconductive solenoid with 1 T magnetic field.

During data taking two independent triggers were used: ``charged'' and
``neutral''. The ``charged'' trigger required at least one track in
the drift chamber while the ``neutral'' one used information about
energies and positions of clusters in the CsI calorimeter.

\section{Event selection}
About $1.1\times 10^5$ collinear events were selected from raw data by
applying the following selection criteria:
\begin{itemize}
\item There is a ``charged'' trigger in event.
\item Two particles with opposite charges were found in the drift chamber.
\item $\rho_{1,2}<0.3$ cm, where $\rho$ is a particle impact parameter
  relative to the beam axis.
\item $|Z_{1,2}|< 7$ cm, where $Z$ is a coordinate at the point of the
  closest approach to the beam line.
\item $|\Delta \phi|=|\pi-|\phi_1-\phi_2|| < 0.15$,
  where $\phi$ is an azimuthal track angle.
\item  $|\Delta \theta|=|\pi-(\theta_1+\theta_2)| < 0.25$, where
  $\theta$ is a polar track angle.
\item Momentum of a final particle, $P$, should be smaller than
  350~\MeVc, and the transverse momentum $P\times\sin{\theta}$ should
  be larger than 90~\MeVc. The last condition ensures a constant
  ``charged'' trigger efficiency. At lower momentum the efficiency
  starts decreasing.
\item The ``average'' polar angle of two particles
  $\theta_\mathrm{aver} = (\pi+\theta_2-\theta_1)/2$ is inside
  $\theta_\mathrm{min} < \theta_\mathrm{aver} < \pi -
  \theta_\mathrm{min}$, where $\theta_\mathrm{min}$ = 1.1 rad. This
  condition decreases a systematic error due to angular uncertainties.
\end{itemize}
A small amount of events has more than two tracks reconstructed in the
tracking system. As a rule, this happens because of mistakes of the track 
reconstruction algorithm  or particles coming back from the calorimeter to
the drift chamber. In this case all pairs of particles with opposite
charges have been checked and an event is accepted if at least one pair
satisfies selection criteria.

\section{Event separation} 
\begin{figure}[tb]
  \includegraphics[height=0.99\linewidth]{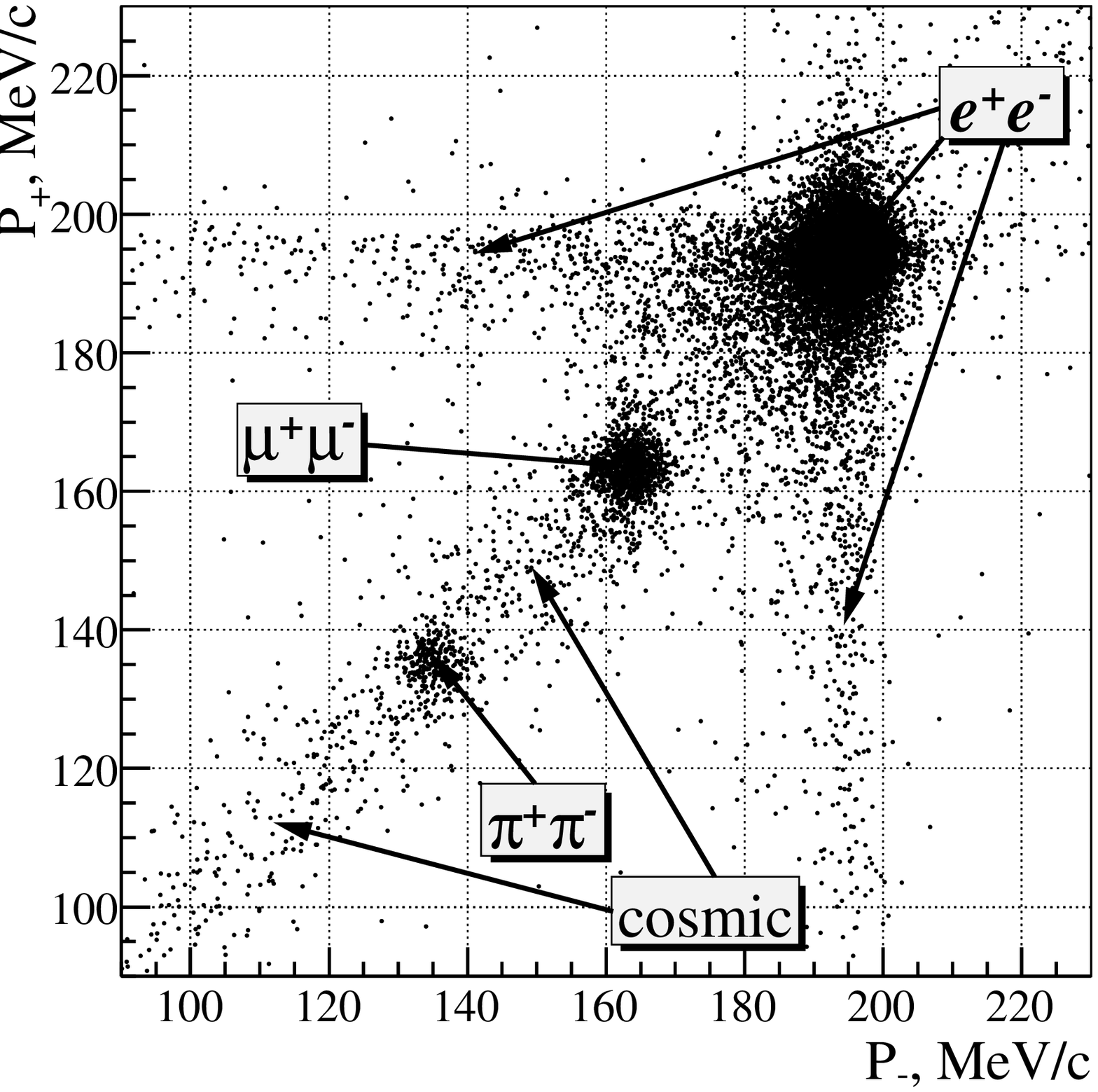}
  \caption{\label{fig:195} Scatter plot $P_{+}$ vs. $P_{-}$ for
    experimental data at the beam energy of 195 MeV.}
\end{figure}

The separation of selected collinear events is based on particle momenta
measured in the tracking system of the detector. A scatter plot
$P_{+}$ vs. $P_{-}$ at the energy
$\sqrt{s}=2\times195$ MeV is shown in Fig.~\ref{fig:195}. One can clearly
see regions where electrons, muons and pions are concentrated as well
as events at the diagonal of the histogram from cosmic 
background.
\begin{figure}[tb]
  \includegraphics[height=0.99\linewidth]{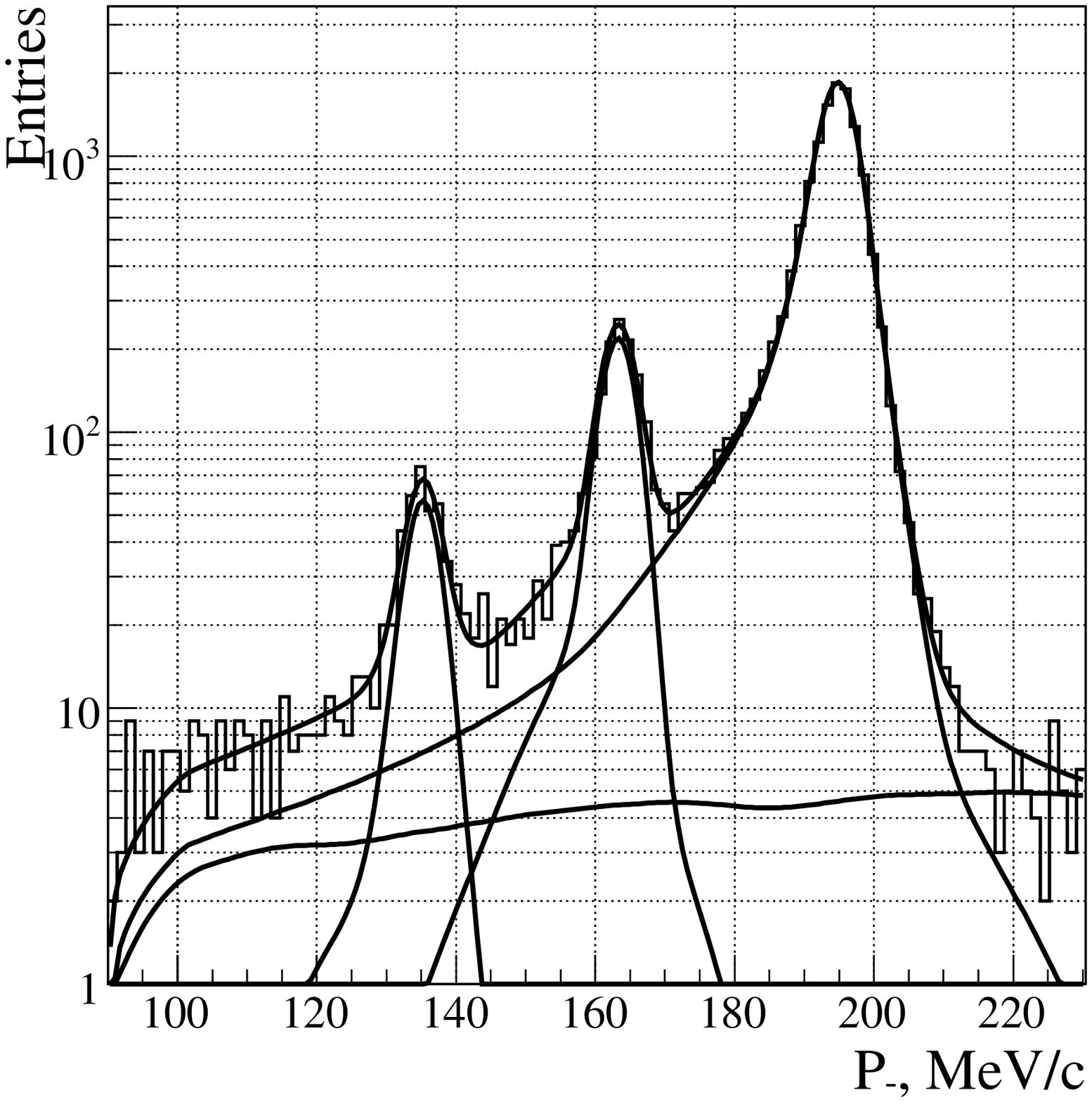}
  \caption{\label{fig:expfit195n} Projection of the two-dimensional
    approximation of the momentum distribution of experimental
    data. Curves show contributions from electrons, muons, pions and
    cosmic background obtained by the event separation procedure.}
\end{figure}
  
The numbers of events of each type are obtained by minimization of the
following likelihood function:
\begin{equation}
  \mathcal{L}=-\sum_\mathrm{events}\ln\Bigl(\sum_{i}w_i\cdot
  f_i(P_{-},P_{+})\Bigr),\ \ \sum_{i}w_i = 1,
  \label{eq:like}
\end{equation}
where $f_i(P_{-},P_{+})$ is the probability density function for final 
particles of the type $i$ which can be electrons, muons, pions or cosmic 
backgound to have momenta $P_{-}$ and $P_{+}$ in the tracking system and
$\displaystyle w_i=N_i/N_\mathrm{tot}$ is the fraction of events of
each type in the total number of events.

As an example, we show in Fig.~\ref{fig:expfit195n} the 
 projection of the two-dimensional approximation onto 
the negative particle
momentum at the the beam energy of 195~MeV.

The selected collinear events consist of 95066 events of the process
\eeee, 9000 events of \eemm, 4053 events of \eepp and 4632 events of
cosmic background.

\section{Pion form factor determination}
The cross section of the process \eepp can be written as:
\begin{equation}
  \sigma^{(B)}_{\pi^+\pi^-}(s)=
  \sigma^\mathrm{(P.L.)}_{\pi^+\pi^-}(s)\cdot|F_\pi(s)|^2,
\end{equation}
where $\sigma^\mathrm{(P.L.)}_{\pi^+\pi^-}(s)$ is the cross section of the
pointlike pion pair production, $F_\pi(s)$ is the pion electromagnetic
form factor.  Note that vacuum polarization effects are included in
the pion form factor definition and for the calculation of the dispersion
integral~(\ref{eq:dispint}) it is necessary to exclude this
contribution and also add the contribution from the process with photon
emission by final pions~\cite{fsr}:
\begin{equation}
  \sigma^{(0)}_{\pi^+\pi^-(\gamma)}(s) =
  \sigma^{(B)}_{\pi^+\pi^-}(s)\cdot\lambda(s)\cdot|1-\Pi(s)|^2,
\end{equation}
where $\lambda(s)$ is the final state radiation correction, $\Pi(s)$ is
the photon polarization operator.

The form factor squared $|F_{\pi}(s)|^2$ is evaluated from experimental
data using the following expression:
\begin{equation}
  |F_{\pi}|^2=\frac{N_{\pi\pi}}{N_{ee}+N_{\mu\mu}}
  \frac{\sigma^{(B)}_{ee}(1+\delta_e)\varepsilon_e(1-
\Delta_\mathrm{B})+\sigma^{(B)}_{\mu\mu}(1+\delta_{\mu})
\varepsilon_{\mu}}
       {\sigma^{(B)}_{\pi\pi}(1+\delta_{\pi})(1+\Delta_{H\&D})
\varepsilon_{\pi}},
\end{equation}
where $N_{ee}$, $N_{\mu\mu}$, $N_{\pi\pi}$ are the numbers of electron,
muons and pions, obtained in the event separation procedure,
$\sigma^{(B)}_{ee}$, $\sigma^{(B)}_{\mu\mu}$, $\sigma^{(B)}_{\pi\pi}$ are
Born cross sections, $\delta_e$, $\delta_{\mu}$, $\delta_{\pi}$ are
radiative corrections at chosen selection criteria, $\Delta_{H\&D}$ is
the correction for pion losses due to nuclear interaction and decays in
flight, $\varepsilon_e$, $\varepsilon_{\mu}$, $\varepsilon_{\pi}$ are
detection efficiencies which depend on the trigger and
reconstruction algorithm, $\Delta_\mathrm{B}$ is the correction for
bremsstrahlung of electrons in the beam pipe and detector material.

The cross section of muon pair production,
$\sigma^{(B)}_{\mu\mu}$, in the first order
of $\alpha$ is known exactly in the frame of QED. By comparing the
theoretical and experimental cross sections one can check the event
separation procedure as well as the calculation of the radiative 
corrections which were used in the cross section determination.

The cross section of the process \eemm is evaluated from experimental data
using the the following expression:
\begin{equation}
  \sigma^\mathrm{exp}_{\mu\mu} =
  \frac{N_{\mu\mu}}{N_{ee}}\frac{\sigma^{(B)}_{ee}(1+\delta_e)
\varepsilon_e(1-\Delta_\mathrm{B})}{(1+\delta_{\mu})\varepsilon_{\mu}},
\end{equation}
where $N_{\mu\mu}/N_{ee}$ is the experimental ratio of the numbers of muons
and electrons obtained in the event separation procedure. For correct 
analysis the ratio of the experimental and theoretical cross sections
$R_{\mu} =
\sigma^\mathrm{exp}_{\mu\mu}/\sigma^\mathrm{theory}_{\mu\mu}$ should
be 1 within experimental uncertainties.

\section{Estimation of systematic errors}  
Various contributions to the systematic uncertainty of the form factor
measurement are shown in Table~\ref{table:syst}. The error caused by
the event separation procedure was estimated from the difference
between the numbers of events in MC and that obtained by the
procedure. The error from the fiducial volume determination was
estimated from the precision of the track end-point measurement in the
Z-chamber. The reconstruction efficiencies were measured separately
for each particle types. The efficiency for electrons was
$96.5\pm0.1$\%, muons -- $96.4\pm0.3$\% and pions -- $97.1\pm0.4$\%.
For the form factor determination of real importance is the difference
between these efficiencies which was taken as an estimate of the
systematic error. Pion losses due to nuclear interactions in the
detector material and decays in flight were determined from the full
simulation of the CMD-2 detector~\cite{cmd2sim}. The accuracy of the
FLUKA package~\cite{FLUKA} for the cross sections of nuclear
interations of low momentum pions was used as an estimate of the
corresponding systematic error. The calculation of the radiative
corrections was based on the work~\cite{radcor} with 0.2\% precision
for each type of processes. The precision of the collider energy
determination was $\Delta E/E \sim 10^{-3}$, resulting in a systematic
error in the pion formfactor of about 0.3\%. Electrons can lose the
large fraction of their energy in the beam pipe and detector material
due to photon emission by bremsstrahlung and thus escape selection
criteria. The precision of this correction depends on the knowledge of
the amount of material with which electrons interact.

\begin{table}
  \caption{\label{table:syst} Contributions to the systematic error.}
  \vspace{0.25cm}
  \centering
  \begin{tabular}{|r|c|}
  \hline
    Source of error & Error, \% \\
    \hline
    Event separation & 0.4 \\
    Fiducial volume & 0.2 \\
    Reconstruction efficiency & 0.2 \\
    Pion loss & 0.2 \\
    Radiative corrections & 0.3 \\
    Energy determination & 0.3 \\
    Bremsstrahlung & 0.05 \\
    \hline
    Total & 0.7 \\
   \hline
  \end{tabular}
\end{table}

\section{Discussion}  
After the event separation procedure and taking into account all
corrections at each energy point, the experimental values of the cross
section of the process \eemm and pion form factor were obtained as
well as the pion cross section for the dispersion
integral~(\ref{eq:dispint}).  These values are shown in
Table~\ref{table:fpi}.

\begin{figure}[tb]
  \includegraphics[height=0.99\linewidth]{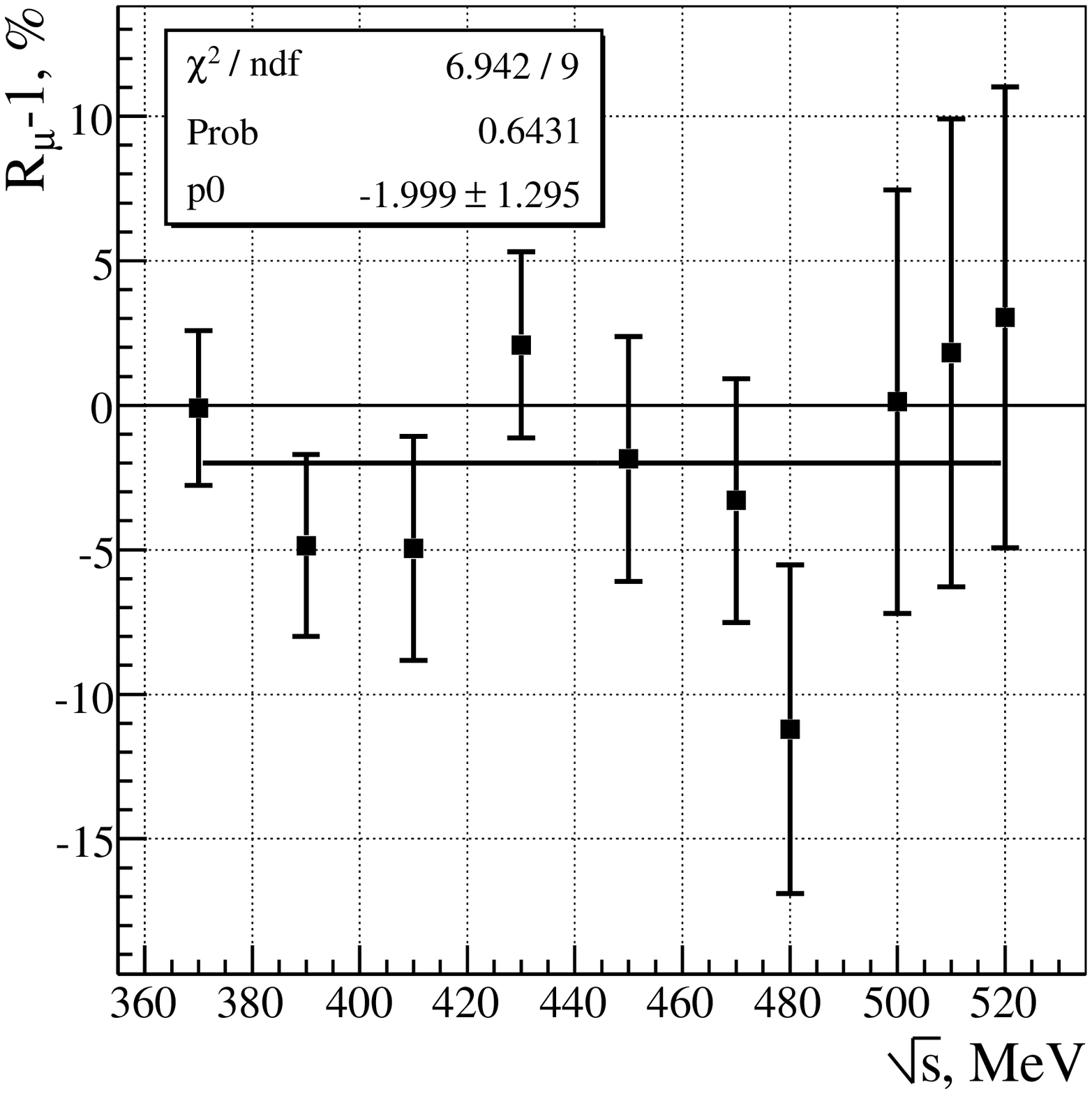}
  \caption{\label{fig:mu_ratio_new} Ratio between experimental and
    theoretical cross sections of muon pair production $R_{\mu}$.}
\end{figure}

The obtained ratio $ R_{\mu} =
\sigma^\mathrm{exp}_{\mu\mu}/\sigma^\mathrm{theory}_{\mu\mu}$ is shown
in Fig.~\ref{fig:mu_ratio_new}. The difference between the
experimental value and theoretical one, averaged over all energies, is
$(2.0 \pm 1.3_\mathrm{stat} \pm 0.7_\mathrm{syst})$~\%.  This is the
first direct comparison of experimental data and theoretical
calculations in the low energy region at the 1\% level.

\begin{table}
  \caption{\label{table:fpi} Cross section of \eemm, electromagnetic
  form factor of charged pion $|F_{\pi}|^2$ and cross section of \eepp
  for the dispersion integral~(\ref{eq:dispint}). Only statistical errors
  are shown.}
  \vspace{0.25cm}
  \centering\begin{tabular}{|c|c|c|c|}
  \hline
  $E$, MeV & 
  $\sigma^\mathrm{(B)}_{\mu\mu}$,  nb &
  $|F_{\pi}|^2$ &
  $\sigma^0_{\pi\pi(\gamma)}$,  nb \\   \hline
  185 &  605 $\pm$ 16 & 2.05 $\pm$ 0.12 &  91.8 $\pm$  5.6\\ \hline 
  195 &  523 $\pm$ 17 & 1.83 $\pm$ 0.12 &  89.0 $\pm$  5.9\\ \hline 
  205 &  476 $\pm$ 19 & 1.98 $\pm$ 0.14 & 100.0 $\pm$  7.0\\ \hline 
  215 &  468 $\pm$ 14 & 2.52 $\pm$ 0.11 & 129.7 $\pm$  5.9\\ \hline 
  225 &  412 $\pm$ 17 & 2.69 $\pm$ 0.15 & 138.5 $\pm$  7.5\\ \hline 
  235 &  373 $\pm$ 16 & 2.83 $\pm$ 0.14 & 144.2 $\pm$  7.3\\ \hline 
  240 &  329 $\pm$ 21 & 3.02 $\pm$ 0.20 & 152.4 $\pm$ 10.3\\ \hline 
  250 &  343 $\pm$ 25 & 3.24 $\pm$ 0.24 & 160.2 $\pm$ 11.7\\ \hline 
  255 &  336 $\pm$ 26 & 3.83 $\pm$ 0.25 & 186.6 $\pm$ 12.4\\ \hline 
  260 &  327 $\pm$ 25 & 3.52 $\pm$ 0.21 & 169.2 $\pm$ 10.3\\ \hline 
  \end{tabular}
\end{table}

For determination of the pion electromagnetic radius $\langle
r_{\pi}^2\rangle$ all CMD-2 data on the pion form factor were
considered~\cite{cmd2ffrho, cmd2ffabovephi}.  To describe
experimental data, the VDM based model from the previous CMD-2 work on
the pion form factor measurement around the $\rho$-meson~\cite{cmd2ffrho} 
was used.

\begin{figure}[tb]
  \includegraphics[height=0.99\linewidth]{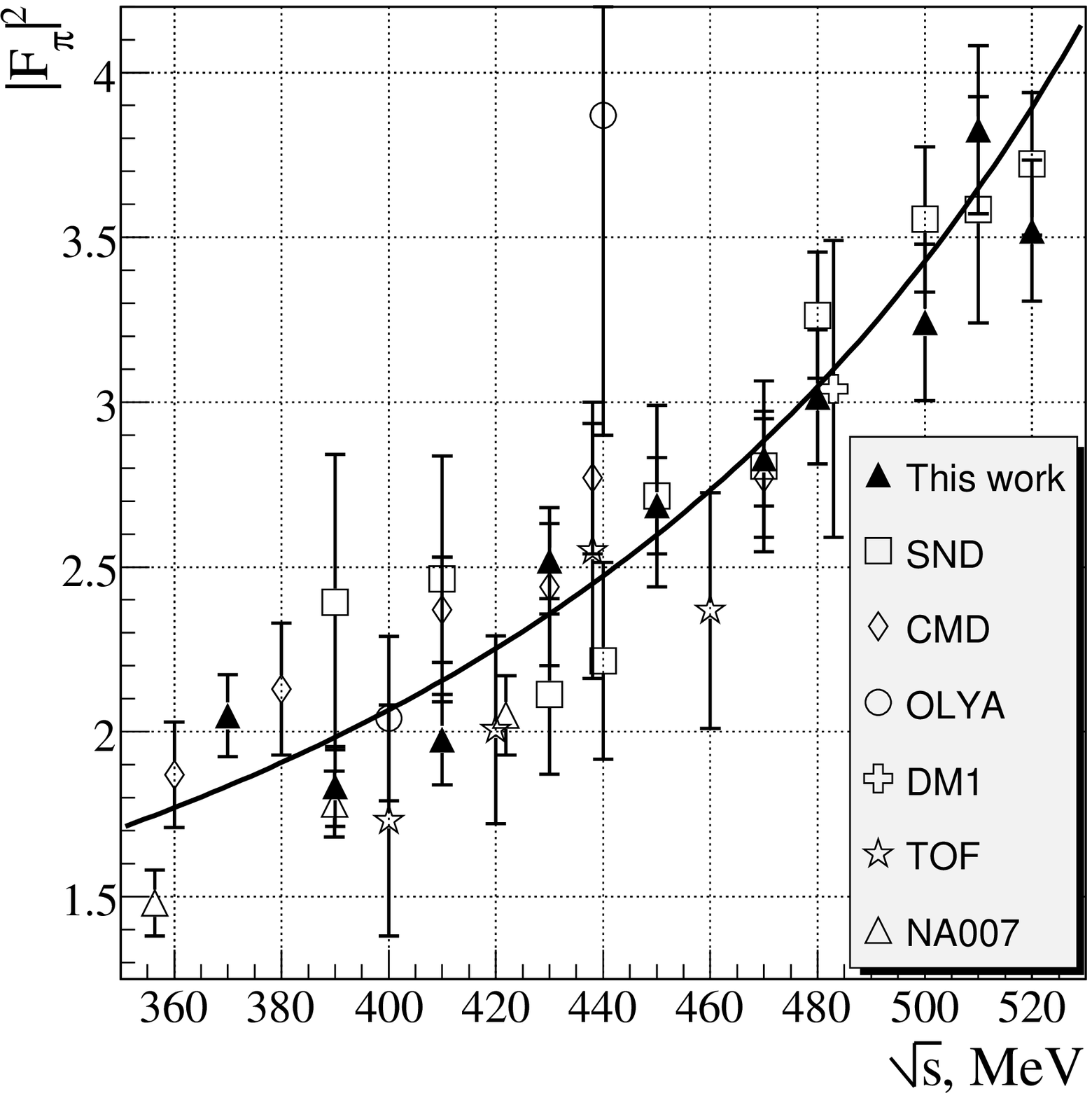}
  \caption{\label{fig:fpifit_low_diff} Comparison with other
  experiments. The curve is the approximation of all CMD-2 pion form
  factor data.}
\end{figure}
The experimental values of the pion form factor measured in this work and
previous experiments~\cite{DM1,TOF,NA71,OLYACMD,sndffrho} are shown in
Fig.~\ref{fig:fpifit_low_diff}.  The electron-positron collider
experiments are in good agreement -- the average difference between
experiments does not exceed 1.5 standard deviations, while the difference
with the NA7 experiment is almost 4 standard deviations.

The pion electromagnetic radius depends on the form factor behaviour at 
low momentum transfers:
$$\langle r^2_{\pi}\rangle = 6\left.\frac{\dd F_{\pi}(s)}{\dd
s}\right|_{s=0}.$$ The value obtained from the form factor approximation
to the point $s=0$ is $$\langle r^2_{\pi}\rangle = 0.4219 \pm 0.0010 \pm
0.0012\ \mbox{fm}^2,$$ where the first error is statistical and the
second one is systematic.  This value is in good agreement with the
result from work~\cite{OLYACMD}, $\langle r^2_{\pi}\rangle = 0.422 \pm
0.003 \pm 0.013\ \mbox{fm}^2$. It is also consistent with the
result from the NA7 experiment in the spacelike region~\cite{NA72}, $\langle
r^2_{\pi}\rangle = 0.439 \pm 0.008\ \mbox{fm}^2$.

The hadronic contribution to ${a}^{\rm had,LO}_{\mu}$ from 
the energy range 390-520~MeV calculated from~(\ref{eq:dispint}) and measured
values of $\sigma^{(0)}_{\pi\pi(\gamma)}$ is
$(46.17\pm0.98\pm0.32)\times10^{-10}$. This contribution agrees
with the value $(48.72\pm1.45\pm1.51)\times10^{-10}$ calculated
from data of previous experiments~\cite{TOF,OLYACMD} and is two times 
more precise.

\section{Conclusion}  
Results of the measurement of the cross section of the process
$e^+e^-\to\pi^+\pi^-$ in the 370-520 MeV c.m. energy range are
presented. A direct test of the QED prediction for the muon pair
production cross section has been performed. The pion electromagnetic
formfactor has been measured with the world best statistical and
systematic accuracy.  Using all CMD-2 pion formfactor data the value
of the pion electromagnetic radius has been calculated.

This work is partially supported by the Russian Foundation for Basic Research,
grants 03-02-16477, 03-02-16280, 04-02-16217,
04-02-16223, 04-02-16434 and 06-02-16156.

\end{document}